\documentclass{elsart}
\usepackage[english]{babel} 
\usepackage{times}
\usepackage{graphicx}
\usepackage{paralist}
\usepackage{cite} 
\begin{document} 
\begin{frontmatter}

\title{Evolution of Strength and Failure of SCC during Early Hydration} 

\author[a1]{Linus K. Mettler},
\author[a1]{Falk K. Wittel\corauthref{cor}}
\corauth[cor]{Corresponding author.}
 \ead{fwittel@ethz.ch}
\author[a1]{Robert J. Flatt}
\author[a1]{Hans J. Herrmann}
\address[a1]{Institute for Building Materials, ETH Zurich, Stefano-Franscini-Platz 3, CH-8093 Zurich, Switzerland.}

 \begin{abstract}
The early strength evolution of self-consolidating concrete (SCC) is studied by a set of non-standard mechanical tests for compressive, tensile, shear and bending failure. The results are applicable in an industrial environment for process control, e.g. of slip casting with adaptive molds in robotic fabrication. A procedure for collapsing data to a master evolution curve is presented that allows to distinguish two regimes in the evolution. In the first, the material is capable of undergoing large localized plastic deformation, as expected from thixotropic yield stress fluids. This is followed by a transition to cohesive frictional material behavior dominated by crack growth. The typical differences in tensile and compressive strength of hardened concrete are observed to originate at the transition. Finally, the evolution of a limit surface in principal stress space is constructed and discussed.
\end{abstract}
\end{frontmatter}

\section{Introduction}
Novel robotic fabrication techniques for non-standard concrete structures rely on the interplay of control systems and the evolution of material performance during setting and early hydration. One example is an innovative process for slip-forming of concrete pillars with variable cross-section using a flexible actuated mold [1]. The mold is significantly smaller than the structures produced and is continuously raised or "slipped" by a robot. The morphological design space is significantly extended with respect to traditional slip casting by actively deforming the hydrating concrete as it leaves the mold. However, the trade-off between deformability and strength restricts the time window for shaping to only a few minutes. The process is constrained by two types of failure: (1) local liquefaction due to the thixotropic nature of the material, leading to collapse of the structure and (2) crack formation due to the rapid rise of the plastic yield stress with respect to the current tensile strength.

The technique has so far mostly employed a polymer fiber reinforced self-consoli-dating concrete (SCC), but has also successively been tested with steel fibers and standard steel reinforcement. The process involves first the preparation of one large batch of heavily retarded SCC which is consecutively accelerated in smaller portions prior to casting [2]. For a review on retarders and superplasticizers as well as their impact on hydration, readers may refer to [3-5]. As the evolution of mechanical performance is affected by various factors (e.g. details of the mixing protocol, variations in raw materials, temperature, humidity, etc.), continuous monitoring by means of simple on-line rheological measurements is necessary at all times during the process. The interpretation of such measurements requires knowledge of the evolution of the failure envelope, i.e. the failure criterion for general multi-axial stress states. Both the fluid and solid properties of concrete have been studied extensively in the context of conventional casting techniques [6]. On one hand studies on the setting and strength development in fresh concrete typically report first measurements at an age of several hours [7]. On the other hand, studies on the workability of concrete focus on the thixotropy of fresh cement pastes typically for time spans up to one hour [8,9]. In this time scale, yield strength is determined first by colloidal interactions and later on by the strength of the first soft, then increasingly rigid percolating network with CSH bonds. Explanatory models based on bonded particles suggest a linear stress-time relation after mixing up to 25 minutes after mixing [9].

However, in studies over longer time periods an exponential relation was observed [10,11]. For the problem described above, it is crucial to describe the rheological and mechanical properties of the material during the early phase of its transition from a non-Newtonian yield stress fluid to a cohesive frictional material.

To assess this early setting regime by mechanical testing for different stress states, "non-standard" mechanical tests were devised covering a wide range of hydration states within the described bounds of liquefaction and fracture. They capture the strength evolution of the initially ductile material to a point where failure is brittle and starts to become dependent on hydrostatic stress. A set of 6 different test setups is described, the respective results are discussed – for simplicity without fiber reinforcement – and condensed into failure envelopes for plane stress evolving with the advancing state of hydration.
\section{Material and methods}
The objective of this study is to determine the evolution of strength for arbitrary stress states in fresh SCC, starting even before the onset of setting. Material handling and test setups are designed to minimize unintentional thixotropic structural breakdown and to withstand gravity early on. All experiments were carried out in a single campaign. This work is limited to a single concrete mixture, designed to meet the requirements for the process of adaptive slip-forming.
\subsection{Material and specimen preparation}
The SCC comprises (in grams per liter of the final mixture) CEM I 52.5 R Portland cement (981.73 g/l), sand aggregates of up to 4 mm in diameter (740.34 g/l), the mineral admixtures fly ash (164.67 g/l) and silica fume (92.89 g/l), as well as a superplasticizer (4 g/l) as a water reducing admixture and water (371 g/l). The SCC is heavily retarded by means of a sucrose solution (30\% D(+)sucrose (99.7\%) and 70\% water, 2.7 g/l) and subsequently accelerated (60 g/l) when taken from the retarded batch, to obtain a constant workability and strength evolution over the entire timespan of the slip-forming process. The slump diameter is typically 20 to 22 cm and the final density of the mixture amounts to 2400 kg/m$^3$.

Special attention was given to maintaining a consistent mixing procedure for all experiments using a rotating pan mixer at predefined mixing speeds (40 or 80 rpm) and intervals (in min) interrupted by cleaning of edges and blades (step denoted cl). After homogenizing the dry material at 80 rpm for 5 min (denoted by 80/5) all fluid parts except the accelerator were added and mixed 40/5 and 80/1. The sequence cl-40/1-cl-80/3 followed, before accelerator was added and finally mixed with 80/5. Between 5 to 10 min after mixing the samples were cast into the respective molds and put to rest. Because of the thixotropic structural buildup at rest, it is crucial to cast all samples within the shortest time span possible (cf. [8-9,12]).
\subsection{Test setup and procedure}
All samples were stored and tested under controlled climatic conditions (50\% relative humidity at 20$^\circ$C) for a period of up to 12 hours after mixing, with experiments starting as early as possible with each test method. The penetration test (Fig.1a) employed a force gauge, recording forces up to 1000 N with 0.2 N accuracy. For all other setups (Fig.1b-f) a universal testing machine, equipped with a 10 kN load cell with 0.1 N accuracy was used. Note that due to the fresh state of the material, an accurate determination of the degree of hydration through calorimetry and other means proved unfeasible. Rate effects, such as thixotropy or viscoelasticity, are not addressed here. Instead, a constant displacement rate $v$ of 1 mm/s was employed, which lies in a regime where the dependence on the test rate was observed to vanish. In addition, the chosen rate allowed for swift execution of each measurement and thus depicted a snapshot in the material evolution where ongoing hydration and flocculation during testing were negligible.

A total of 6 different mechanical test setups were employed, named (a)-(f) and shown in Fig.1a-f with dimensions being summarized in Tab.1. Specimen size effects, e.g. arising from the presence of aggregates, were precluded by initial tests with varying dimensions. Each test series of (b)-(f) was accompanied by penetration tests (a) on the same material. The specimen geometries and loading conditions are described in the following.
\begin{figure}[htb] \centering{ \includegraphics[width=14.cm]{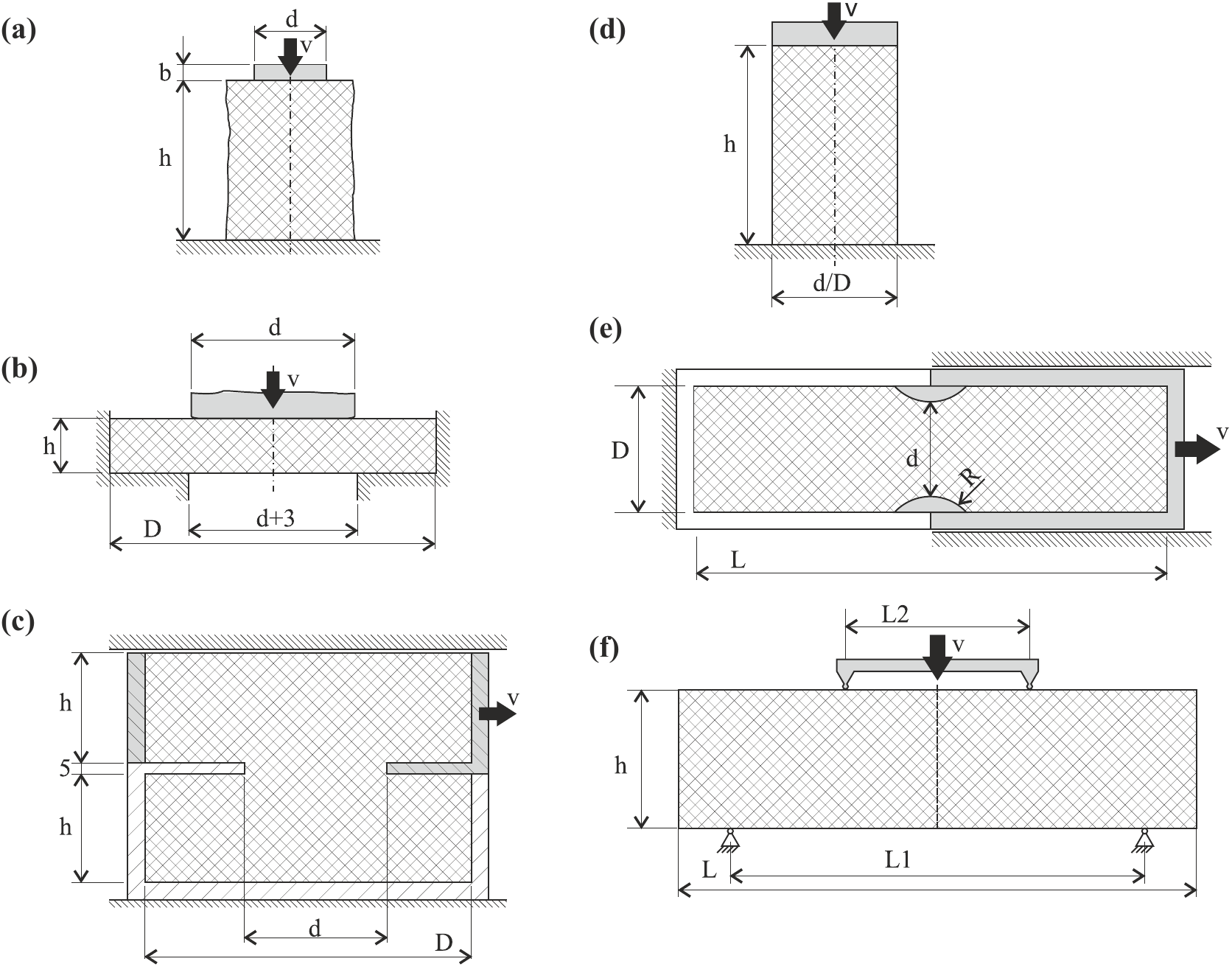}
 \caption{(a) Penetration of a round disk into a body of concrete, (b) punching of a circular hole into a concrete layer, (c) shearing of an interface between two blocks of concrete, (d) compression of a cylindrical sample, (e) extension of a notched bar, and (f) four-point bending of a beam with square cross-section. Dimensions see Tab.1. Moving parts are shaded in gray.} }
\end{figure}
\paragraph*{Test (a) - Penetration:} A rigid cylinder of diameter $d$ and height $b$ was driven at constant velocity $v$ into a basin of concrete of depth $h$. The resistance experienced by the indenter was recorded as a function of the penetration depth. The total displacement imposed was of the order of $d$ which is sufficient to reach a steady state flow for fresh concrete. The test was fully automatized using a tri-axial robot and 3 concrete basins, each accommodating 17 separate penetration measurement points. The penetration test forms the basis of the subsequent analysis. It is performed on all material samples in parallel to the other tests (b)-(f). The test is based on ASTM C403/C403M-08 [13] and its use for this application has already been reported in [14].
\paragraph*{Test (b) - Punch-through:} A cylindrical indenter with diameter $d$ cuts a hole into a flat cylindrical concrete sample at constant $v$. The associated reaction force was recorded versus depth. The sample of diameter $D$ and height $h$ was vertically and radially supported by a mold comprising a hole of diameter slightly larger than $d$. The sample is thin in the sense that $h\ll d\pi$, yet thick enough to contain a sufficient amount of aggregates to assume a homogeneous material. This setup is inspired by the standard shear punch test, e.g. ASTM D732-10 [15].
\paragraph*{Test (c) - Shear:} The setup for measuring early shear strength is similar to the Jenike shear cell (ASTM D6128-14 [16]). The upper compartment (gray hatched area in Fig.1c) was horizontally displaced at constant rate $v$, measuring its resistance. A steel cable translated the vertical movement of the testing machine via a pulley to the sample. Contrary to a classical shear cell, no normal force was applied and two horizontal plates, attached to the lower and upper compartments respectively, produced a notch for cleaner shear conditions at the fracture plane. An array of pins near the lower and upper surfaces of the sample (not shown in Fig.1c) ensured a rigid body motion of the two concrete blocks even when the sample was fresh. Note that rollers at the top suppressed the separation of the compartments.
\paragraph*{Test (d) - Compression:} Cylindrical concrete samples were vertically compressed by a plate moving at rate $v$. The sample diameter $d$ was chosen large enough to avoid size effects due to inhomogeneous particle distribution at walls, and $h/d$ was big enough to allow a slip layer to form diagonally to the loading direction. Samples were cast on a base plate using pipe segments lined with a thin sheet of Teflon treated with a mold-releasing agent. Just before testing the sample was demolded and the Teflon sheet carefully detached to preserve the thixotropic structural buildup in the sample at early stages of hydration.
\paragraph*{Test (e) - Tension:} Tensile testing was implemented using a mold of two separable parts. Unlike conventional tensile specimens, fresh concrete cannot be clamped. In order to enhance shear force transmission from the mold to the sample, several rows of pins were attached laterally to the box walls, decreasing in size towards the center of the specimen. A round notch of radius $R$ reduced the effective cross section of the sample to $A_{eff}=dh$ with width $d$ and height $h$. Note that the notch segments were free to move after casting and therefore did not exert any significant loads on the material during testing. At time of testing the right half (gray in Fig.1e) was horizontally separated from its fixed counterpart (white) at constant velocity $v$ using a similar cable-and-pulley configuration as in test (c).
\paragraph*{Test (f) - Bending:} To obtain the flexural strength, four-point bending tests on prismatic samples of size $L\cdot b\cdot h$ similar to ASTM C78/C78M [17] were performed as soon as the material was able to withstand the gravitational load at the supports (distance $L_1$). The loading frame (distance $L_2$) moved at v while the reaction force of the sample was recorded. Tests are valid only if a tensile crack formed at the lower side of the sample within $L_2$, where the bending moment is maximal and homogeneous. Note that contrary to tests (a)-(e) four-point bending was only applicable to specimens from a relatively advanced state of hydration on.
\begin{table*}[htb]\label{tab1}
\caption{Summary of dimensions of samples in tests (a)-(f) (cf. Fig.1). Dimensions marked with * are not shown in Fig.1.}
  \begin{tabular}{p{0.7cm}p{0.7cm}p{1.7cm}p{0.7cm}p{0.7cm}p{1cm}p{3.5cm}p{1cm}} \hline
	test & $b$ [mm]&$d$ [mm]&$h$ [mm]&$D$ [mm]& $L/L_{1,2}$ [mm]&$A_{eff}$ [mm$^2$]& $N$ \\
	(a)	& 4	& 18.8	& 40	& -	& -	& $(d/2)^2\cdot \pi$	& $>$500 \\
	(b)	& -	& 60	& 20	& 120	& -	& $dh \pi$	& 50 \\
	(c)	& 60$^*$	& 60	& 50	& 150	& -	& $bd$	& 41 \\
	(d)	& -	& 63	& 100	& 63	& -	& $(d/2)^2\cdot \pi$	& 93 \\
	(e)	& -	& 60 (R=31)	& 60$^*$	& 80	& 300	& $dh$	& 43 \\
	(f)	& 60$^*$	& -	& 60	& -	& 225 / 180 / 80	& $2/3(bh^2)/(L_1-L_2)$	& 16 \\ \hline	
  \end{tabular}
\end{table*}
\section{Experimental Observations}
For each sample the resistance force was recorded as a function of displacement. Additionally, the deformation patterns were observed visually in order to distinguish between "plastic" and "brittle" failure, when possible. 
\subsection{Force-Displacement Curves}
Force-displacement curves measured at different age since mixing, like the examples shown in Fig.2 for each test at different hydration states, are the basis for obtaining the strength evolution in time. As typical for such tests, the increasing slope in the first millimeters of each test is due to the displacement until initial contact with the specimen is established (tests (a),(b),(d)) or until horizontal load cables are pre-tensioned (tests (c),(e)). Large plastic deformations at the contact zones of fresh samples in bending test (f) lead to an overestimation of the deflection. The resulting compliance of the system superimposes the elastic regime of the material, thus impeding stiffness measurements, but has no effect on the magnitude of the maximum force.
\begin{figure}[htb] \centering{ \includegraphics[width=14.cm]{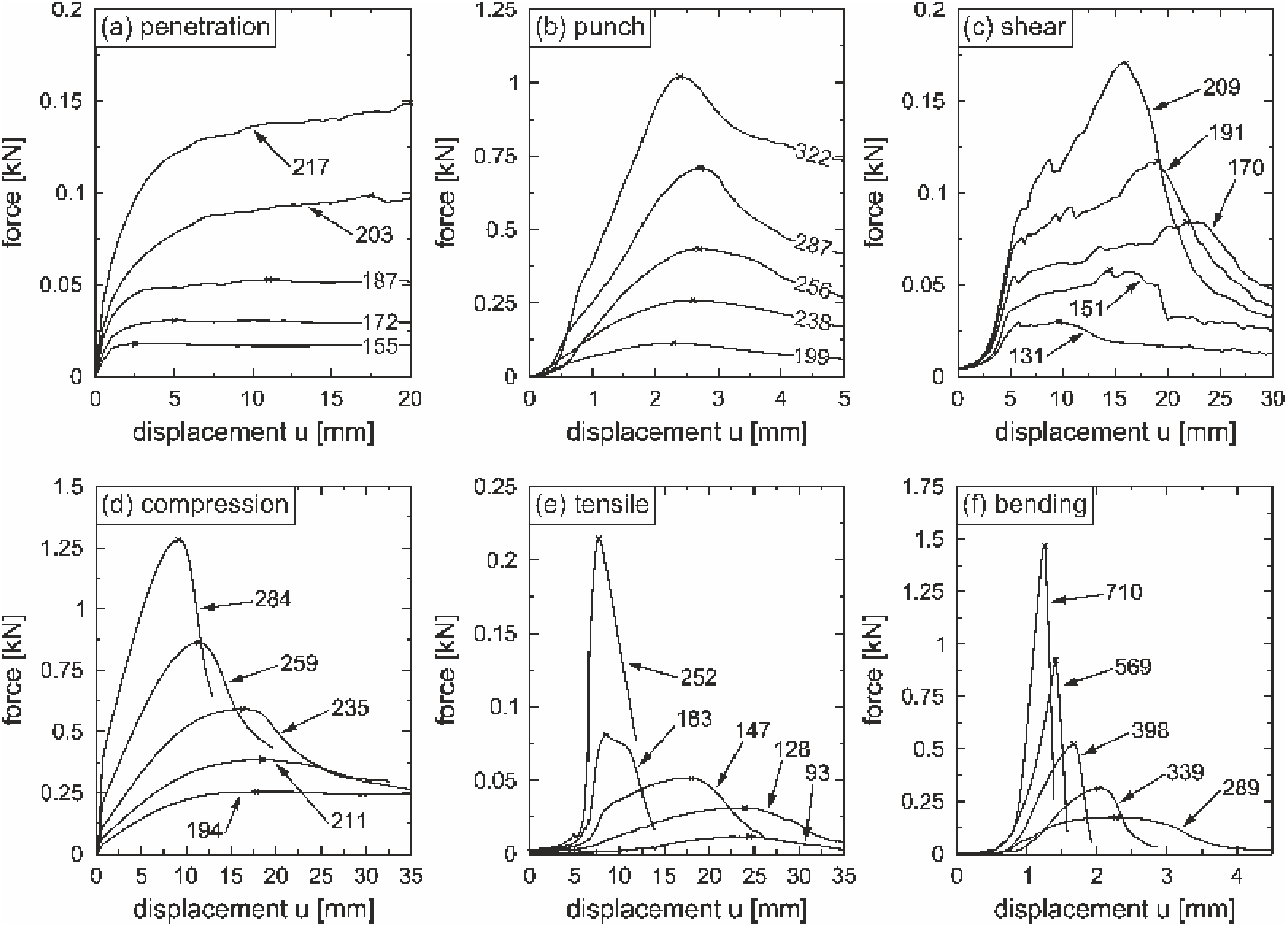}
 \caption{Exemplary force-displacement curves at different degrees of hydration for (a) penetration, (b) punch, (c) shear, (d) compression, (e) tensile and (f) bending tests. The specimen age is indicated for each curve in minutes after mixing.} }
\end{figure}
\subsection{Failure Patterns}
Qualitative observations of the failure evolution and emerging patterns give an important insight into the materials behavior for the different setups and times after mixing. The penetration test (a) exhibits an increasing difficulty to reach steady state flow (i.e. constant resistance force at constant velocity) as hydration progresses (see Fig.2a). Instead, the force tends to increase with penetration depth as expected for solid-like behavior to a point where cracks suddenly propagate radially from the edge of the indentation through the basin. At this point, the test method ceases to provide data independent from the geometry of the boundary of the basin. Cross-sections of penetration samples made of colored concrete layers, see Fig.3, illustrate the deformation pattern of fresh concrete reaching steady state. As the indenter penetrates, an undisturbed concrete "plug" of conical shape outlined by a thin region of high shear forms in front of the disk, increasing the effective surface of the indenter. A steady shear flow is induced around the plug, provided that the material is fresh enough to flow.
\begin{figure}[htb] \centering{ \includegraphics[width=8.cm]{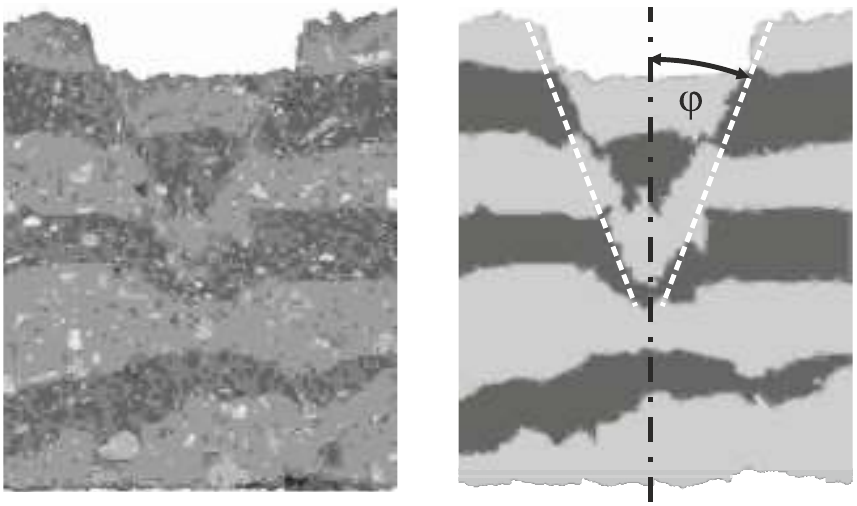}
 \caption{Compression cone in a penetration experiment 170 min after the end of mixing. Left: photograph of a cut section, right: same section after image processing.} }
\end{figure}
The punch and shear tests (b), (c) induce a homogeneous shear-dominated stress state along an interface inside the sample, inducing local shear failure. However, at a distance from this interface the shear stress decreases significantly. Small cracks are confined to the proximity of the interface even if they wanted to extend further into the bulk of the material. It is therefore possible that an alleged shear crack in fact consists of a connection of many small diagonal tensile cracks. Furthermore, the presence of aggregates at the sheared interface prevents the formation of a clean crack, thus contributing to the dissipation of energy.

At early times, cylinder compression tests (d) exhibit a shear band formation in a plane of maximum shear, i.e. at 45$^\circ$ to the loading direction (Fig.4a), dividing the sample into two relatively undeformed parts. The measured force reaches a maximum as the band is fully formed and remains nearly constant as the upper part of the specimen slides along the diagonal shear plane. Hence very young samples exhibit steady state flow at a localized shear band similar to the penetration test. As the material becomes increasingly solid, the specimen accommodate the deformation more globally. A multitude of vertical cracks propagate through the sample as a result of circumferential tensile stresses arising from the increase in cross section (see Fig.4b). The peak force roughly coincides with the formation of such cracks, which dramatically lower the remaining strength of the sample thereafter. A plug of conical shape is found in front of the indenter. This failure pattern is qualitatively very similar to fully hydrated concrete samples under uniaxial compression [18].
\begin{figure}[htb] \centering{ \includegraphics[width=12.cm]{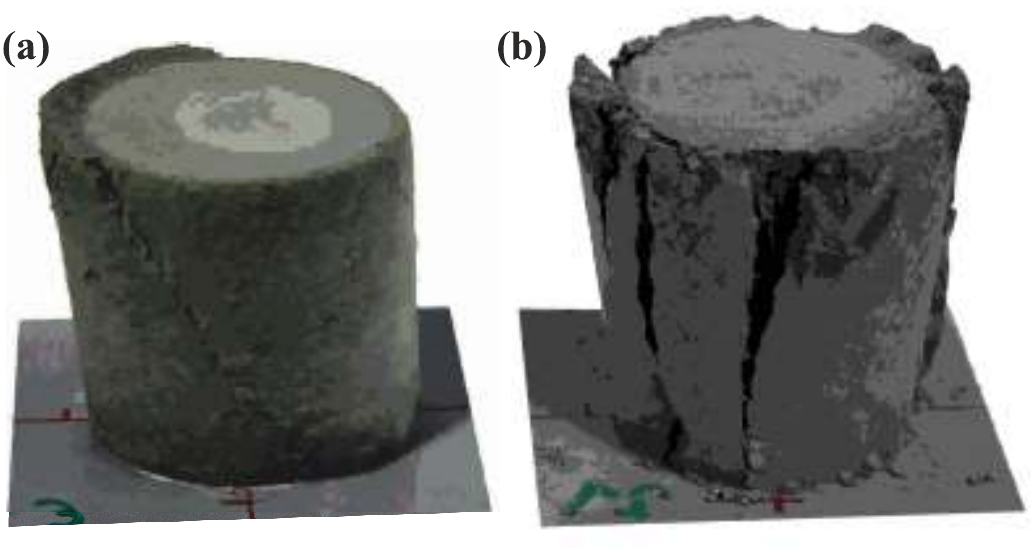}
 \caption{Failure patterns at early ((a) 140 min) and later ((b) 260 min) stages of hydration.} }
\end{figure}
Similar to compression, the tensile test (e) of relatively fresh samples shows a tendency to form a slip layer at 45$^\circ$C to the loading direction (see Fig.5a). This is consistent with the pronounced softening branch in Fig.2e even without fiber reinforcement. However, the mechanism of failure evolves quickly towards a single crack orthogonal to the loading direction, separating the sample (Fig.5b). Accordingly, the decline in measured force is abrupt (Fig.2e). The behavior of the bending test (f) is similar (Fig.2f), although this test is applicable only at later times due to high localized stresses at the contact zones.
\begin{figure}[htb] \centering{ \includegraphics[width=14.cm]{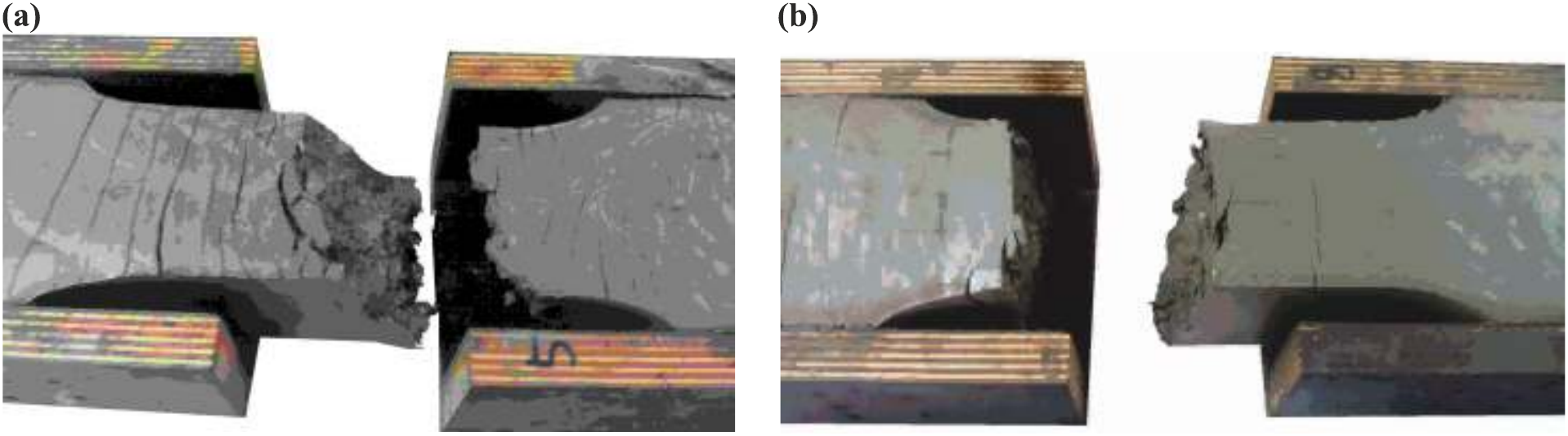} }
 \caption{Crack morphology at early ((a) 160 min) and later ((b) 240 min) stages of hydration.} 
\end{figure}
\section{Data Processing}
A comprehensive constitutive model ranging from the early state of the yield fluid to the solidified state of the frictional cohesive material is missing up to date. Without a priori knowledge of a suitable constitutive model, the definition of material parameters such as yield stress or strength is ambiguous [19]. The choice of experimentally identified parameters originates from the demand for finding a predictive model for the collapse of structures fabricated by adaptive slip-forming. In this section the derivation of strength evolution based on the measured data is described. The procedure includes a practicable definition of strength, a correction for gravity effects and a correction for inconsistencies in the evolution of different batches of material.
\subsection{Strength}
The key property is the strength for multi-axial stress states and its evolution. Strength is defined for each test setup, at a certain time with respect to the end of the mixing procedure, as the maximum measured resistance force normalized by the initial reference area of the sample. For practical purposes the rheological behavior is neglected by fixing the displacement rate at a value that minimizes rate dependencies. Note that heat release due to hydration plays a minor role, since the time frame of hydration in this study is within the "period of slow reaction" [20] and hence is not accessible by calorimetry. 

The reference area corresponding to each test is listed in Tab.1. It intuitively represents the cross-section of the sample suffering the highest homogeneous stress. However, the choice is not always clear, as in case of the penetrometer test (a) [21]. For this test, penetration forces are normalized by the bearing area of the disk as suggested in [15], despite its tendency to form a cone-shaped plug. The bending test (f) does not exhibit a homogeneous stress state either. Instead, the flexural strength is defined as the maximum tensile stress along the beam cross-section. It is expected to relate qualitatively to the tensile strength of the material, being slightly larger than the tensile strength.
\subsection{Corrections for Gravity}
Testing starts as soon as possible without the samples collapsing prematurely. The gravitational loads are thus of the same order as the early strength and should be accounted for in the interpretation of results, in particular for the compression and bending tests (d), (f). Failing to do so would imply a substantial underestimation of early strength. In detail, compressive strength calculations include an additional gravitational load of $\rho gh$. The flexural strength or “modulus of rupture” is calculated as $MoR=1.5F_{max}\cdot (L_1-L_2)/(bh^2)+0.75\rho g L_1^2/h$, where the second term incorporates the contribution of gravitational load to the bending moment.
\subsection{Reference Time Correction Method}
When summarizing all strength measurement points versus time since mixing in one plot, it becomes evident that despite of extraordinary diligence, significant deviations occur across experiments carried out with different batches of material in the same campaign. Fig.6a illustrates the extent of scatter in penetration and compression test data. These inconsistencies are thought to be mainly due to variations in the raw materials as well as environmental changes over the course of the test campaign, e.g. temperature and humidity during mixing. For a meaningful interpretation these effects need to be corrected for, while preserving the validity of the results.

The correction procedure is based on the observation that despite the overall fluctuation, penetration data (test (a)) from within one mixture (i.e. measured at different locations of the same basin of concrete) follows approximately an exponential evolution law between 10 kPa and 1 MPa. Consequently, each penetration series/batch $n$ can be fitted with an exponential curve (e.g. dashed line in Fig.6a). Equivalently an exponential "reference curve" can be constructed from an average of all penetration test series (solid line). Each measurement point $i$ of series $n$ is compared to the reference curve and shifted in time as illustrated in Fig.6a (inset):
\begin{itemize}
\item Consider the measurement $(t_i^n,S_i^n)$. According to the exponential fit to series $n$ (dashed line), the estimated penetration strength at time $t_i^n$ is $S_{p,i}^n$.
\item The reference penetration curve reaches the same strength $S_{p,i}^n$ at a different time $t_i^{ref}$.
\item Hence the data point is shifted by $\Delta t_i$ to the new location $(t_i^{ref}, S_i^n)$ in order to compensate for the difference in evolution of set $n$.
\item The routine is not only applicable to the correction of penetration data, but is used for all test setups (a)-(f) by considering any measurement $(t_i^n, S_i^n)$ in connection with the corresponding penetration strength $S_{p,i}^n$ of the same series $n$.
\end{itemize}

A mixture does not always evolve like the reference, but its current state can be determined from the reference data simply by a penetration measurement that reveals the current state $t^{ref}$. The corrected strength data are shown in Fig.6b with respect to the reference time $t^{ref}$. Note that while deviations across different series of experiments are greatly reduced, the method preserves scatter inherent to the respective test setup as well as systematic deviations from the exponential fit. 
\begin{figure}[htb] \centering{ \includegraphics[width=14.cm]{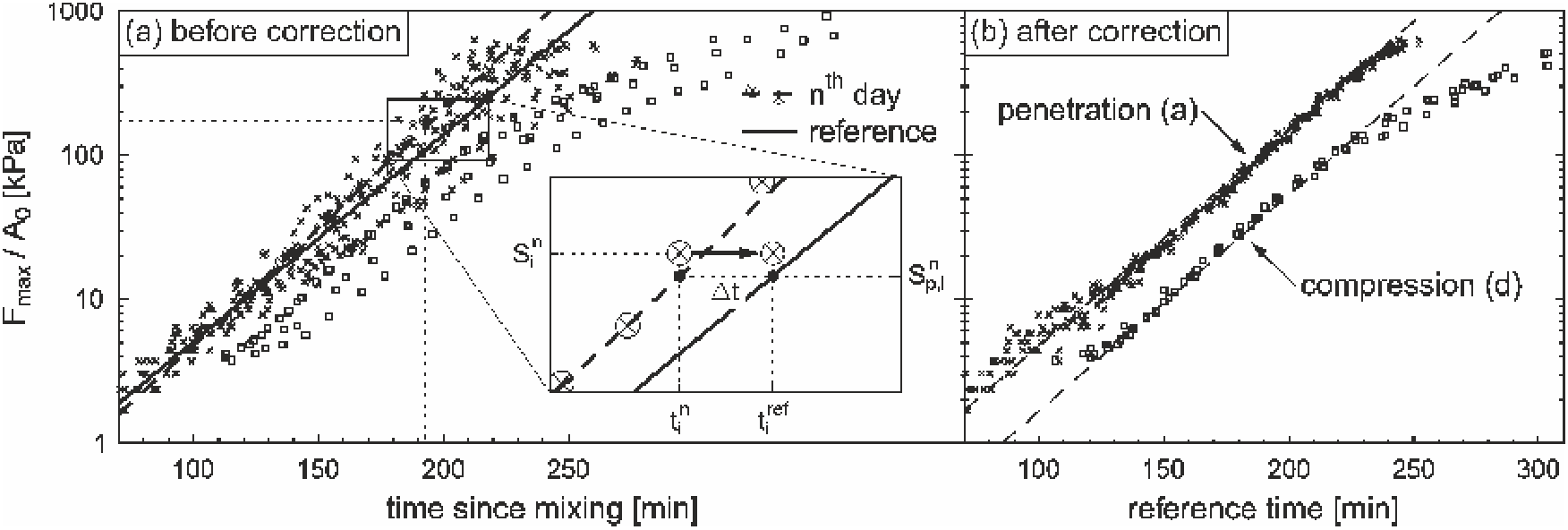} }
 \caption{Penetration (crosses) and compression (squares) strength data (a) before and (b) after application of the time shift method to correct for inconsistent strength evolution across mixtures prepared on different occasions. Inset: Illustration of the time shift of strength data point $i$ of experiment set $n$ $(t_i^n, S_i^n)$ towards a reference penetration curve. The method assumes that penetration strength evolves exponentially between 10 kPa and 1 MPa.} 
\end{figure}
\section{Discussion of merged results}
Finally, results from the different setups need to be combined into one coherent framework for strength evolution of the material. Its sensitivity to the environment necessitates a large set of experiments along with the described correction method. Only after obtaining data collapse, the strength evolution with the transition from the initially ductile failure at a shear band to the prevalence of cracks becomes observable.
\subsection{Evolution of Strength}
The compression (test (d)) and penetration (test (a)) strength data in Fig.6b initially grow exponentially in good approximation, expressed by the relation $S=a\exp(bt)$, with different prefactors $a$ but similar exponent $b$ (as indicated by parallel lines). However, with progressing hydration the compressive strength departs from the exponential path to a significantly slower growth rate. This apparent change of regime coincides with the transition of the samples’ failure pattern from a localized shear zone to distributed cracks (as depicted in Fig.4).

The corrected data of all tests is summarized in Fig.7. Distinct scaling regimes are particularly pronounced when comparing compressive with tensile tests (e), (f) (Fig.7a). The initially fresh samples exhibit largely the same strength in both uniaxial compression and tension, i.e. their strength appears to be independent of the hydrostatic component of stress. Thereafter the growth in tensile strength slows down dramatically as cracks appear. The change in failure mode of the compression test (d) occurs 60 minutes later. The estimated transition times for all tests are listed in Tab.2. In the following the two regimes are discussed separately.
\begin{figure}[htb] \centering{ \includegraphics[width=14.cm]{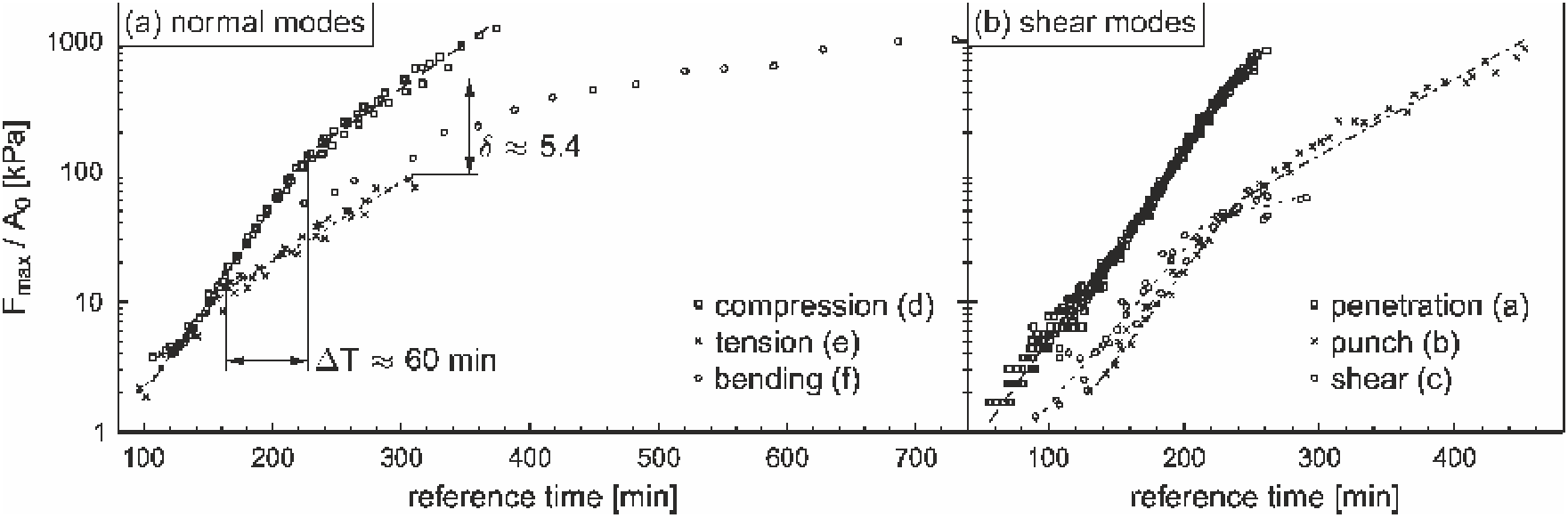} }
 \caption{(a) Comparison of compressive, tensile and bending strength evolution. Until 3 hours after mixing the measured strength is independent of the loading direction. Thereafter, the compressive strength exceeds the tensile strength by $a$ factor $\delta$. The bending strength ($MoR$) suggests a further slowdown in strength increase with advancing hydration. (b) Comparison of penetration, punch and shear strength evolution. The regime transitions in punch and shear appear to occur at similar times as in compression.} 
\end{figure}
\begin{table*}[htb]\label{tab1}
\caption{Estimated transition from 1$^{st}$ to 2$^{nd}$ regime.}
  \begin{tabular}{p{5cm}p{3cm}p{3cm}} \hline
	  & time [min]& stress [kPa] \\
		(test (b)) Punch & 250.2 & 81.6 \\
		(test (c)) Shear & 219.4 & 44.1 \\
		(test (d)) Compression & 227.4 & 137.5 \\
		(test (c)) Tension & 167.5 & 13.1 \\ \hline	
  \end{tabular}
\end{table*}
\begin{table*}[htb]\label{tab1}
\caption{Fitting parameters with 95\% confidence intervals of 1$^{st}$ regime with coefficient of determination $R^2$ and sample size $N$.}
  \begin{tabular}{p{3.7cm}p{3.7cm}p{2.7cm}p{0.7cm}p{0.7cm}} \hline
	& $a$ [kPa] &	$b$ [min$^{-1}$] &	$R^2$ &	$N$ \\
(test (a)) Penetration &	0.188 [-0.011,+0.011]	&0.033 [$\pm$0.0003]	&0.992	&310\\
(test (b)) Punch	&0.035 [-0.010,+0.015]	&0.031 [$\pm$0.002]	&0.982	&23\\
(test (c)) Shear	&0.098 [-0.036,+0.056]	&0.028 [$\pm$0.003]	&0.942	&28\\
(test (d)) Compression	&0.070 [-0.009,+0.010] &	0.033 [$\pm$0.001]	&0.993	&56\\
(test (e)) Tension	&0.161 [-0.075,+0.138] 	&0.026 [$\pm$0.005]	&0.927	&14 \\	\hline	
  \end{tabular}
\end{table*}
\paragraph*{First "Plastic" Regime:}  The exponential fit parameters of the first regime are summarized in Tab.3. Again, the similarity of exponents $b$ and the disparity in the prefactors $a$ is evident. This conceptually corresponds to a scaling between the curves, where the scaling factor of each test is closely related to the different stress states in the samples and the choice of reference area $A_0$ used for the conversion of force to strength (Tab.1). With the knowledge of the three-dimensional stress state of each test it should thus be possible to deduce the shape of the yield surface from the scaling factors. Moreover, the similarity in exponents $b$ in Tab.3 suggests that the yield surface just expands equally in all directions in the first regime at an exponential rate.

As a consequence, the first regime of tests (a)-(e) (Fig.8a) can be collapsed to a single evolution curve of "equivalent uniaxial strength" (Fig.8b), when data of each test is scaled with the following constant conversion factors. The punch/shear strengths are increased by a factor of $\sqrt{3}$  according to the von Mises equivalent stress under pure shear [22]. This assumption is supported by the ductile nature of the material in this regime and the equivalence of tensile and compressive strength data. For the penetration test neither the natural choice of $A_0$ nor the stress state are trivial. Instead the penetration strength is scaled to the effective surface area of the plug cone estimated from Fig.3, thus decreased by the factor $A_{cone}/A_0=(\sin \phi)^{-1}$. The compression and tensile strength remain unscaled. Note that pure shear conditions are assumed for the punch test, thereby neglecting the additional compressive stress component on the sheared surface caused by the radial confinement of the sample. Consequently, the "equivalent strength" derived from the punch tests is consistently underestimated (cf. Fig.8b), indicating that a significant confining pressure is prevalent in the test.
\begin{figure}[htb] \centering{ \includegraphics[width=14.cm]{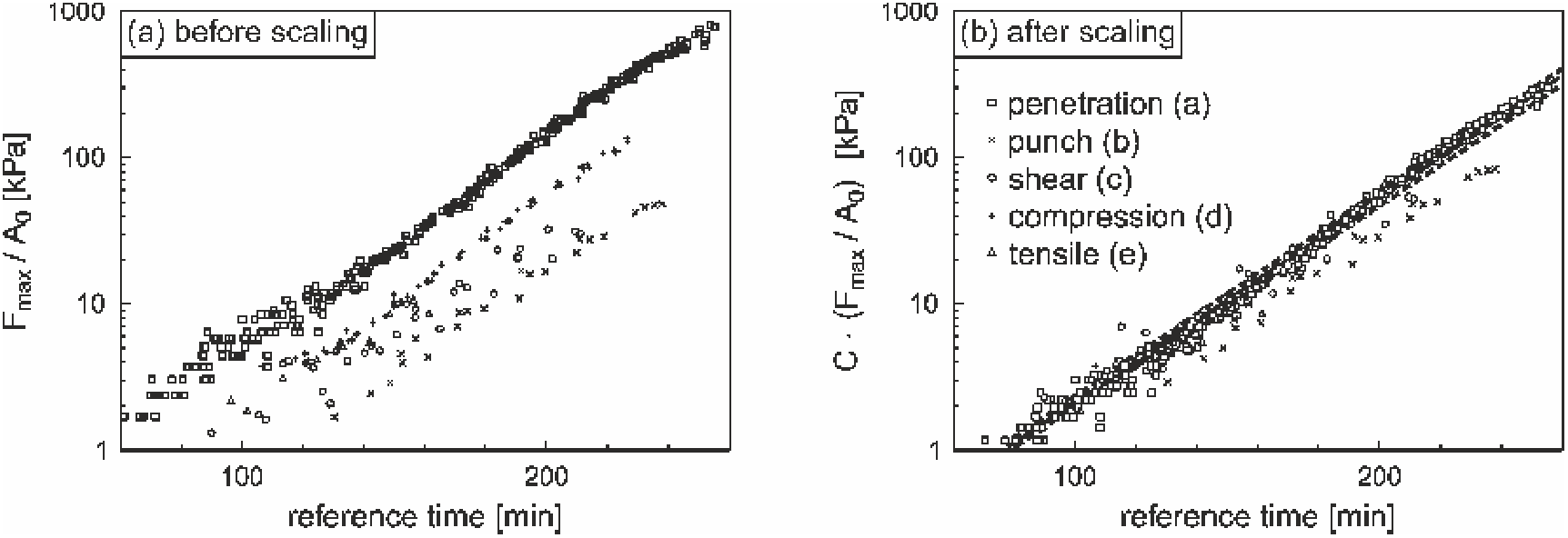} }
 \caption{(a) Strength evolution of the first regime, i.e. where deformations localize in a shear plane. (b) Evolution of equivalent strength estimated from tests (a)-(e) by accounting for the penetration cone size or by assuming Mises plasticity. All data more or less collapses to a single curve with parameters $a=0.0745 [-0.004;+0.004]$ kPa and $b=0.0328 [-0.0003;+0.0004]$ min$^{-1}$ for 95\% confidence interval (dashed lines). } 
\end{figure}
\paragraph*{Second "Brittle" Regime:}The strength gain in the second regime, albeit slower, can again be approximated by an exponential function for about one order of magnitude of strength in the observed time span (Fig.7 and coefficients in Tab.4). The exponents of the evolution law for the punch, compression and tension tests are similar, implying that in this regime the ratio between compressive and tensile strength remains roughly constant (at 5.4 in this case). The shear strength exhibits a significant amount of scatter due to frictional forces in the test setup and thus remains inconclusive. Of course a further slowdown in strength gain is anticipated for all tests until the concrete is fully hydrated and strength converges to a final value. This trend is indicated by the four-point bending test data (see Fig.7a), which is qualitatively related to the tensile strength.

It is important to observe that a very large portion of the well-known difference in compressive and tensile strength of hardened concrete emerges already at the early transition from a pressure-insensitive thixotropic yield stress fluid to a brittle cohesive frictional solid. It is not a putative difference in rate (or exponent) of compressive and tensile strength growth that gives rise to the pressure sensitivity of the final material, but rather their different times of transition from the fluid to the solid regime. The difference in “brittle” compression and tension may always be existent, except that in fresh concrete the low plastic yield strength undercuts these failure modes. The solid regime thus emerges only after the plastic strength, growing at a higher rate, surpasses brittle strength – a transition which occurs at different times for compression and tension because of the inherent difference in brittle compressive and tensile strength.
\begin{table*}[htb]\label{tab1}
\caption{Fitting parameters with 95\% confidence intervals of 2$^{nd}$ regime with coefficient of determination $R^2$ and sample size $N$.}
  \begin{tabular}{p{3.7cm}p{3.7cm}p{2.7cm}p{0.7cm}p{0.7cm}} \hline
	& $a$ [kPa] &	$b$ [min$^{-1}$] &	$R^2$ &	$N$ \\
(test (b)) Punch	& 4.0 [-1.08,+1.48]	0.012 &[$\pm$0.001]	& 0.967 	&27\\
(test (c)) Shear&	13.9 [-9.33,+28.29]	0.005& [$\pm$0.0044]	& 0.381	&13\\
(test (d)) Compression	&3.7 [-0.91,+1.22]	&0.016 [$\pm$0.001]	& 0.967	&37\\
(test (e)) Tension&	1.3 [-0.28,+0.36] &	0.014 [$\pm$0.001] 	& 0.963	&29\\ \hline	
  \end{tabular}
\end{table*}
\subsection{Strength Envelope}
Finally, the strength envelope, generally represented by a convex surface in 6-dimensional stress space, is constructed. The envelope joins all points of failure in the stress space, thus delimiting the field of stable stress configurations within the envelope from the failure field outside the envelope. Owing to the isotropic behavior of the (unreinforced) concrete and the ensuing rotational symmetries of the elasticity tensor, the strength envelope can be expressed in only 3 dimensions, e.g. in terms of the principal stresses. Each strength measurement represents a point on the strength envelope for the given state of hydration (or reference time since mixing).

The above test setups, with the exception of the penetration and punch tests for which the tri-axial stress state is not a priori known, impose plane stress conditions where at least one principal stress component is negligible. Therefore, it is sufficient to examine failure in the $\sigma_1-\sigma_2$-plane, where the strength envelope reduces to a closed convex curve. Such curves can be estimated for a given time after mixing by taking strength values from the exponential fits to tensile, compressive and shear test data as shown in Fig.9 for 100 to 300 min after mixing. Curves are drawn through these points in the $\sigma_1-\sigma_2$-plane by means of a spline interpolation in cylindrical coordinates ${r,\phi}$. The early strength data, up to an order of a few kPa, correspond well with the von Mises criterion. Thereafter strength becomes increasingly pressure sensitive, thus the curves extend far into the pressure zone. Clearly, the accuracy of the resulting envelope could be greatly improved by collecting more experimental data under various bi-axial stress states.
\begin{figure}[htb] \centering{ \includegraphics[width=14.cm]{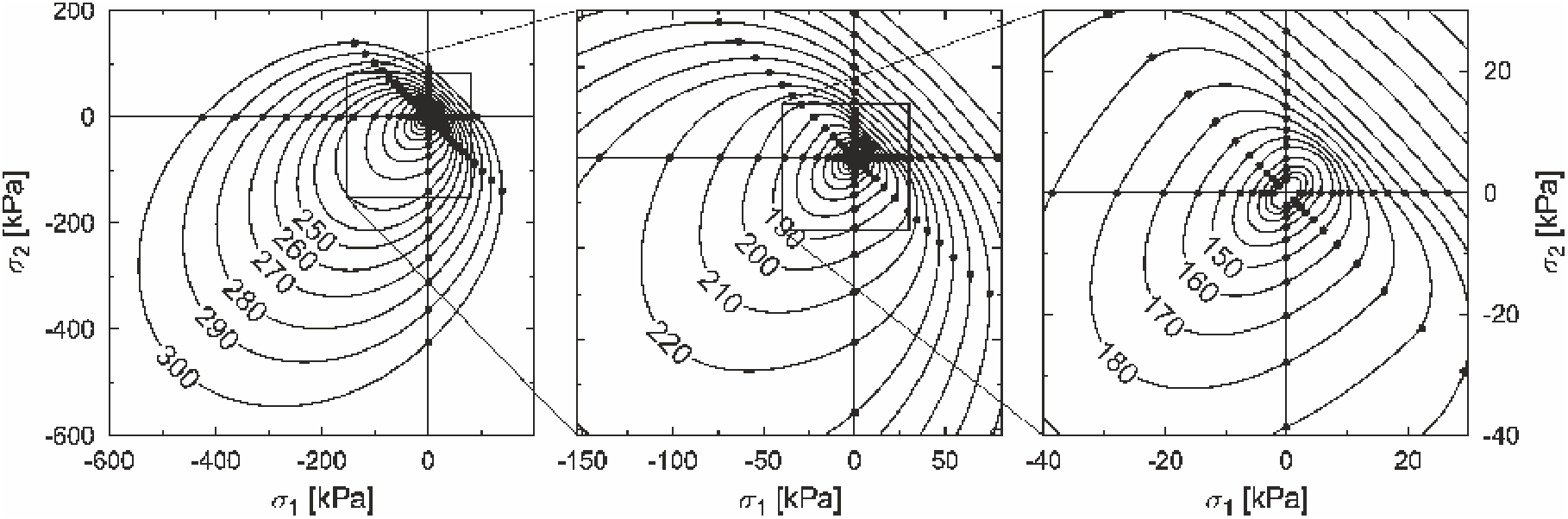} }
 \caption{Rough indication of time dependent failure surfaces in plane stress from 100 to 300 min after mixing, based on the exponential fits to compression, tension and shear strength measurements. The numbers denote the reference time in minutes. The early curves (100-140 min) resemble the Mises equivalent stress, but they quickly evolve into pressure-sensitive behavior due to the different times of regime change of each test setup.} 
\end{figure}
\subsection{Friction Angle}
Additionally, it is possible to calculate the cohesion $c$ and internal friction angle $\phi$ based on the compressive and tensile strength evolution e.g. for the Coulomb-Mohr and Drucker-Prager models. The Drucker-Prager yield surface describes an infinite cone in principal stress space, where $c$ is the intercept of the cone with the principal stress axes and $2\phi$ describes the opening angle (aperture) of the cone. The cohesion, a measure of overall strength particularly under tension, increases roughly exponentially in the observed time frame. As expected, $c$ continues to grow at a slower rate after the regime transition. The friction angle, closely related to the difference in compressive and tensile strength, increases rapidly from 0 to 48 degrees at the transition, and is set to grow only slightly thereafter.
\begin{figure}[htb] \centering{ \includegraphics[width=7.cm]{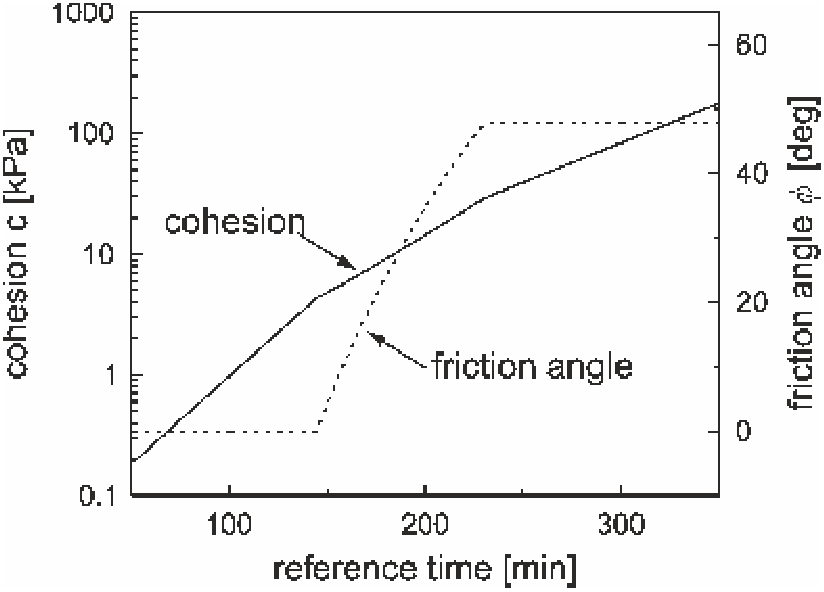} }
 \caption{Cohesion and friction angle as a function of reference time since mixing. The cohesion grows roughly exponentially, while the friction angle jumps from 0 to 48 degrees during the transition.} 
\end{figure}
\section{Summary and Conclusions}
The strength evolution during early hydration is of increasing importance with the growing recognition for new technologies in concrete fabrication such as concrete printing or adaptive slip casting. It was shown by evaluating more than 500 individual samples in six different non-standard experimental setups at times of up to 12 hours after mixing, that two principal regimes govern the strength evolution: A first one, where the material is ductile and where shear, compressive and tensile strengths are well-represented by the von Mises criterion. At the transition, typically around 3 hours after mixing for the SCC studied, cracks begin to prevail and failure is increasingly dependent on the hydrostatic component of stress, as seen e.g. by the divergence between compressive and tensile strengths. The second regime is characterized by significantly slower strength increase, but at a similar rate for tension and compression. In both regimes, strength scales exponentially with time over several orders of magnitude. This differs with respect to observations on small time scales accessible to rheometers, where a linear relation is found [9]. To make the findings applicable to more general states of stress, failure envelopes in plane stress space at different concrete ages were constructed, which would finally converge to the final strength envelope known for hardened high-strength concrete, reported for biaxial stress states e.g. in [23-24]. Despite variations in the strength evolution of different material batches, the current failure envelope is accessible by a simple penetration measurement. 

There are limitations to this study that go beyond the choice of a single concrete mix. Considerable differences in evolution occurred across samples due to variations in raw materials and temperature. A meaningful interpretation of the collected data necessitated correction procedures, such as the time shift method proposed herein. By shifting solely in time (and not in force), however, it is assumed that the penetration force is a state variable uniquely describing the material. Thus it is ignored for now, that flocculation and hydration can evolve differently – therefore calling for at least two state variables and evolution laws [25] – and that the same strength does not necessarily imply an identical state of the material. Instead, samples were casted systematically at the same time after mixing and it was assumed that flocculation and hydration depend equally on temperature (or that the effect of temperature on flocculation is very small compared to hydration). However, the influence of flocculation on the yield stress is limited to time scales that are significantly below the ones addressed in this study [9]. The validity of these assumptions was demonstrated only within the parameter space of the experiments that keeps in mind the technological design window of the adaptive slip-casting process.
\section*{Acknowledgements}
The support by the ETH Zurich under ETHIIRA grant no. ETH-13 12-1 "Smart Dynamic Casting", as well as from the European Research Council advanced grant no. 319968-FlowCCS is acknowledged. We thank David Walker, Lukas Bodenmann and Stefan Aebersold, who contributed to this work by preparatory studies and also express our gratitude to Ena Lloret, Nicolas Roussel and Peter Fischer for numerous useful discussions on the topic.
\section*{References}
\begin{itemize}
\setlength{\itemsep}{12pt}
\item[(1)] E. Lloret, A.R. Shahab, L.K. Mettler, R.J. Flatt, F. Gramazio, M. Kohler, S. Langenberg, Complex concrete structures Merging existing casting techniques with digital fabrication, Comput. Aided Design 60 (2014) 40–49.
\item[(2)]	A.R. Shahab, E. Lloret, P. Fischer, F. Gramazio, M. Kohler, R.J. Flatt, Smart dynamic casting or how to exploit the liquid to solid transition in cementitious materials, extended abstract, 7th International RILEM Conference on SCC, Paris (2013).
\item[(3)]	G. Gelardi, S. Mantellato, D. Marchon, M. Palacios, A. Eberhardt, R.J. Flatt, Chapter 9 – Chemistry of chemical admixtures, in Science and Technology of Concrete Admixtures, Ed. P.-C. Aïtcin and R.J. Flatt, Woodhead Publishing (2016), 149-218.
\item[(4)]	D. Marchon, R.J. Flatt, Chapter 8 – Mechanisms of cement hydration, in Science and Technology of Concrete Admixtures, Ed. P.-C. Aïtcin and R.J. Flatt, Woodhead Publishing (2016), 129-145.
\item[(5)]	D. Marchon, R.J. Flatt, Chapter 12 – Impact of chemical admixtures on cement hydration, in Science and Technology of Concrete Admixtures, Ed. P.-C. Aïtcin and R.J. Flatt, Woodhead Publishing (2016), 279-304.
\item[(6)]	K. Kovler, N. Roussel, Properties of fresh and hardened concrete, Cem. Concr. Res. 41 (2011) 775-792.
\item[(7)]	R.C.A. Pinto, A.K. Schindler: Unified modeling of setting and strength development, Cem. Concr. Res. 40 (2010) 58-65.
\item[(8)]	N. Roussel, A thixotropy model for fresh fluid concretes: Theory, validation and applications, Cem. Concr. Res. 36 (2006) 1797-1806.
\item[(9)]	N. Roussel, G. Ovarlez, S. Garrault, C. Brumaud: The origins of thixotropy of fresh cement pastes, Cem. Concr. Res. 42 (2012) 148-157.
\item[(10)] A. Perrot, A. Pierre, S. Vitaloni, V. Picandet : Prediction of lateral form pressure exerted by concrete at low casting rates, Mater. Struct. 48 (2015) 2315-2322.
\item[(11)] K.V. Subramaniam, X. Wang: An investigation on microstructure evolution in cement paste through setting using ultrasonic and rheological measurements, Cem. Concr. Res. 40 (2010) 33-44.
\item[(12)]	G. Ovarlez, X. Chateau, Influence of shear stress applied during flow stoppage and rest period on the mechanical properties of thixotropic suspensions, Phys. Rev. E 77 (2008) 061403.
\item[(13)]	Standard Test Method for Time of Setting of Concrete Mixtures by Penetration Resistance, ASTM C403-/C403M-08 (2008).
\item[(14)]	E. Lloret, L.K. Mettler, A.R. Shahab, F. Gramazio, M. Kohler, R.J. Flatt, Smart Dynamic Casting: A robotic fabrication system for complex structures, 1st Concrete Innovation Conference, Oslo (2014).
\item[(15)]	Standard Test Method for Shear Strength of Plastics by Punch Tool, ASTM D732-10 (2010).
\item[(16)]	Standard Test Method for Shear Testing of Bulk Solids Using the Jenike Shear Cell, ASTM D6128-14 (2014).
\item[(17)]	Standard Test Method for Flexural Strength of Concrete (using simple beam with third-point loading), ASTM C78-/C78M-10 (2013).
\item[(18)]	J.G.M. van Mier: Fracture Processes of Concrete, CRC Press, 1997.
\item[(19)]	P.C.F. Moller, J. Mewis, D. Bonn, Yield stress and thixotropy: on the difficulty of measuring yield stress in practice, Soft Matter 2 (2006) 274-283.
\item[(20)]	J.W. Bullard, H.M. Jennings, R.A. Livingston, A. Nornat, G.W. Scherer, J.S. Schweitzer, K.L. Scrivener, J.J. Thomas, Mechanisms of cement hydration, Cem. Concr. Res. 41 (2011) 1208-1223.
\item[(21)]	D. Lootens, P. Jousset, L. Martinie, N. Roussel, R.J. Flatt, Yield stress during setting of cement pastes from penetration tests, Cem. Concr. Res. 39 (2009) 401-408.
\item[(22)]	J. Lubliner, Plasticity theory (Dover books on engineering), Dover Publications (2008).
\item[(23)]	H. Kupfer, Das Verhalten des Betons unter zweiachsiger Beanspruchung (in German), report, Lehrstuhl für Massivbau, Technical University Munich, no. 78 (1969) 124 ff.
\item[(24)]	A. Nimura, Experimental research on failure criteria of ultra-high strength concrete under biaxial stress (in Japanese), Architectural Institute of Japan, Structures II vol. C (1991) 473-474.
\item[(25)]	P. Coussot, Rheometry of Pastes, suspensions, and Granular Materials: Applications in Industry and Environment, J. Whiley \& Sons Inc., Hoboken, New Jersey (2005).
\end{itemize}
\end{document}